\documentclass[a4paper,12pt]{article}
\usepackage{overcite}
\usepackage{amsmath}
\usepackage{amssymb}
\usepackage{here}

\newcommand{\fett}[1]{\mbox{\boldmath$#1$}}

\newcommand{\allStates}{n_1,\ldots,n_L}
\newcommand{\ONvec}{\left|n_1 n_2 \ldots n_L\right\rangle}
     
\newcommand{\ONstring}{n_1\ldots n_L}

\title{New Approaches for \textit{ab initio} Calculations of Molecules with Strong Electron Correlation}

\author{Stefan Knecht\thanks{Corresponding Author; email: stefan.knecht@phys.chem.ethz.ch},
{Erik Donovan Hedeg{\aa}rd},
{Sebastian Keller},
{Arseny Kovyrshin},\\
{Yingjin Ma},
{Andrea Muolo},
{Christopher J.\ Stein}, and
Markus Reiher\thanks{Corresponding Author; email: markus.reiher@phys.chem.ethz.ch}
}

\begin{document}

\maketitle

\begin{center}
{ETH Z\"urich, Laboratorium f{\"u}r Physikalische Chemie, Vladimir-Prelog-Weg 2,\\ 8093 Z\"urich, Switzerland}\\[1.5ex]

\end{center}

\begin{abstract}
Reliable quantum chemical methods for the description of molecules with dense-lying frontier orbitals
are needed in the context of many chemical compounds and reactions. Here, we review developments that led to our new 
computational toolbox which implements the quantum chemical density matrix renormalization group in a second-generation
algorithm. We present an overview of the different components of this toolbox.

\end{abstract}

\section{Introduction}\label{sec:intro}

Computational modeling has undoubtedly become an integral part of chemical research 
\cite{doi:10.1021/cr2004663}. For instance, understanding a (photo-)chemical process in atomistic detail
--- including all elementary reaction steps involved --- calls for a reliable, but feasible and 
preferably black-box computational approach 
that provides a sufficiently accurate approximation to the exact solution of the electronic Schr{\"o}dinger equation. 
A typical example for a complex chemical process is a reaction catalyzed by a transition-metal complex \cite{negishi2011}. 
A metal center can directly activate a reagent through bond formation and/or bond breaking or act as a 
photoacceptor for a subsequent energy transfer to the reagent from an electronically excited state. 
The common challenge is then to quantitatively describe open shells that will emerge in such processes, which is far from trivial to 
meet \cite{reiher09_tm_rev,cramer_09_dft_tm_review,Boguslawski2011,perspectives_dft12,dmrg-spin-densities12}. 
The presence of open shells and/or activated bonds usually entails a multiconfigurational electronic structure 
where strong static electron correlation becomes sizable. Molecules with these features then exhibit many dense-lying orbitals in
the frontier-orbital region.  

Although by construction a single-configuration \textit{ansatz}, to date, density functional theory (DFT) \cite{perspectives_dft12}\ is by far 
the most popular approach to study photochemistry (see for example Ref.~\citenum{lindh_photochemistry_review2012}\ 
and references therein) and transition-metal chemistry \cite{cramer_09_dft_tm_review,Garino20120134,comp_tools_tm_catalysis_chapter2014} 
because of its low computational cost, often at reasonable accuracy, and because of its favorable scaling with the size of a molecule. 
The standard approach to describe half-filled shells in molecules with small HOMO-LUMO gaps is to break a symmetry of
the system within 
DFT, usually the total-spin symmetry \cite{Noodleman1981,Noodleman1982}.
Besides known fundamental problems of single-configuration DFT in correctly describing strong static correction \cite{failure_of_dft_08_rev} 
all spin symmetries can be properly introduced in DFT leading to spin-state dependent functionals \cite{Jacob2012}, but little
work along these lines has been carried out so far \cite{bogu13}.

A multiconfigurational \textit{wave function} based method tailored to recover static correlation 
is the complete active space (CAS) \textit{ansatz} \cite{QUA:QUA23107}, often combined with a simultaneous 
self-consistent field (SCF) optimization of the orbital basis. Naturally, CASSCF-type approaches have been applied in theoretical photochemistry 
and transition-metal chemistry \cite{pier03,caspt2-review-wires2012,molcas8}. 
The central idea of a CASSCF-type approach is the selection of an active space of $N$\ electrons in $L$\ orbitals to yield a CAS($N$,$L$),
on which a full configuration interaction (FCI) expansion 
\begin{equation}
\label{eq:CASCI}
\left|\Psi_{\rm CAS-CI}\right\rangle  = \sum_{\allStates} C_{n_1 n_2 \dots n_L} \ONvec\ ,
\end{equation}
of the wave function is constructed
($\ONvec$\ is an occupation number vector corresponding to an orthonormal basis state, e.g., a Slater determinant).
This procedure, however, does not yield an exact wave function as the orbital basis is restricted.
Moreover, the underlying FCI expansion still scales exponentially with respect to the number of active electrons and orbitals so that 
the computational feasibility of traditional CAS methods reaches a limit at a CAS size of about CAS(18,18)\cite{molcas8}. 

Originally developed to study the physics of spin chains, 
the density matrix renormalization group (DMRG) \cite{White1992, White1993} algorithm emerged 
as a viable alternative to traditional CAS methods. This is rooted in the fact that it is capable of iteratively converging to the exact solution 
in a given active orbital space with polynomial rather than exponential cost \cite{Schollwock2011}. In DMRG, CASs are 
accessible with up to about 100 orbitals exceeding by far the limits encountered by traditional CAS methods.
Many quantum-chemical DMRG implementations, with (DMRG-SCF) and without (DMRG-CI) orbital optimization,
have been developed since the late 1990s.

Here, we present an overview of our recent efforts to further develop methods that rely on the DMRG.
They comprise features (i) to locate for a given molecule the minimum structure of its 
ground and excited states, (ii) to take into account electron correlation effects beyond static correlation, (iii) to model a complex molecular system embedded in a structured environment, and (iv) to account for effects of Einstein's theory of special relativity, when needed. 
To this end, we developed the \textit{second-generation} quantum-chemical DMRG program {\textsc{QCMaquis}} 
\cite{keller2015efficient}\ that unites these objectives in a unique framework. 
In actual applications, the break-even point for computational costs of a DMRG calculation compared to a traditional 
CASSCF calculation is reached for CAS(14,14) in {\textsc{QCMaquis}}.

%

\section{Second-generation DMRG}\label{sec:secondDMRG}

DMRG\cite{White1992, White1993}
invented by White was inspired by its predecessor, Wilson's numerical renormalization group, 
in which a Hilbert space is truncated by selecting the lowest-lying eigenstates of a Hamiltonian.
By contrast, White proposed to select states according to their weight in the density matrix, which dramatically improved
the performance of the renormalization group. To date, DMRG is the most successful numerical method to solve one- or quasi-one-dimensional systems
in solid state physics. In quantum chemistry, DMRG has been established as a powerful active-space method, allowing
for much larger active spaces than a traditional CASSCF-type approach.
The higher performance comes at a price, however. The truncation of the density matrix by only retaining the $m$ states 
with the highest weight implies that the accuracy of the approximation is to be assessed \emph{a posteriori}.
The latter can be achieved by analyzing the weights of the discarded density matrix eigenstates and by performing
calculations with different values for $m$ (that may then be subjected to an extrapolation towards $m\rightarrow\infty$).

A few years after the introduction of DMRG, it was realized, that the DMRG algorithm is equivalent to the variational optimization of a special class
of \textit{ansatz} states called matrix product states (MPSs) \cite{Oestlund1995, Verstraete2004b}, 
\begin{eqnarray}\label{eq:MPS}
\left|\Psi_{\rm MPS}\right\rangle & =& \sum_{\allStates} M^{n_1} M^{n_2} \cdots M^{n_L} \ONvec\ ,
\end{eqnarray}
where $L$ is the total number of active orbitals as before and $M^{n_l}$\ are matrices for the $l$-th spatial orbital 
whose product yields the corresponding CI coefficients $C_{\ONstring}$ (implying that $M^{n_1}$ and $M^{n_L}$ are actually vectors). 
Each local space $n_l$\ of the $l$-th spatial orbital is of dimension four\ corresponding to the possible orbital occupations $n_l = \left|\uparrow\downarrow\right>, \left|\uparrow\right>, \left|\downarrow\right>, \left|0\right>$.  
The connection between DMRG and MPS provided the theoretical understanding of why DMRG works well for one-dimensional systems 
but becomes less efficient in higher dimensions \cite{Vidal2003, Barthel2006, Schuch2008}. 
Moreover, it allowed for very flexible implementations in which wave functions and operators can be combined arbitrarily in operations such 
as overlap and expectation value calculations, operator-wave function actions, and operator-operator actions.

In quantum chemical MPS-DMRG in conjuction with a Hamiltonian expressed as a matrix product operator (MPO), the main challenge is the
efficient construction of the MPO, because the performance of the method depends critically on it.
We demonstrated\cite{keller2015efficient} that the full quantum-chemical Hamiltonian MPO can be efficiently constructed so that the same computational
scaling is achieved as in traditional, i.e., pre-MPO quantum-chemical DMRG.
We refer to this MPO-based algorithm implemented in {\textsc{QCMaquis}} as second-generation DMRG \cite{keller2015efficient}.

Compared to traditional DMRG, second-generation DMRG is more versatile with respect to the decisive quantities, i.e., wave functions and operators, 
that can be handled independently of each other. As a consequence, for example, we were able to quickly implement relativistic 
Hamiltonians (see Section \ref{sec:relH})
by simply exchanging the MPO while re-using all contraction routines handling the application of the MPO to the MPS.
Another example for the efficiency of a second-generation algorithm is the 
implementation of spin-adapted MPSs and MPOs\cite{keller2016}.

\section{Tensor Network Parameterizations}\label{sec:tns}

The MPS \textit{ansatz} imposes a one-dimensional {\it ad hoc} ordering of molecular orbitals 
in the construction process of the total basis states and wave function. While for linear spin chains in 
solid-state physics this is a natural procedure, for chemical systems this is in general 
not the case and can give rise to convergence problems. To overcome these issues stimulated further 
developments and led to the formulation of a new family of states, the so-called
Tensor Network States (TNS)\cite{nish2000,nish2001,gend2002,gend2003}. The TNS approach tends to break down the high-dimensional 
CI coefficient tensor $C_{n_1 n_2 \dots n_L}$ of the FCI  
\textit{ansatz} into a network of low-rank tensors. One of the latest developments in this field are Tree Tensor Network 
States (TTNS)\cite{shi2006,murg2010,barc2011}. For the quantum chemical 
Hamiltonian a TTNS variant has been developed by Nakatani and Chan\cite{naka2013} that 
is essentially a generalization of the MPS concept. Interestingly, the Nakatani-Chan implementation was an MPO-based second-generation quantum-chemical DMRG program prior to {\textsc{QCMaquis}} with the correct scaling \cite{sharma14}, and has been turned into the general MPO/MPS library MPSXX available on GitHub\cite{chan16}.

In the Nakatani-Chan approach, tensors are connected 
as a tree graph of degree $z$ (any orbital has at most $z$ neighbors) and depth $\Delta$ 
(number of edges/arcs from the root node to the leaf node of the tree). For a given 
orbital (site) $i$ their TTNS \textit{ansatz} reads
\begin{equation}
\left| \Psi_{\rm TTNS} \right\rangle = \sum_{b_i^1 \ldots b_i^z n_i}
C_{b_i^1 \ldots b_i^z}^{n_i} \left| b_i^1 \ldots b_i^z n_i \right\rangle,
\end{equation}
where $\left| b_i^{\alpha} \right\rangle$ ($ \alpha = 1,\ldots, z$) is the renormalized 
basis in the $\alpha$-th branch of site $n_i$\cite{naka2013}. Nakatani and Chan  define this basis recursively 
contracting tensors in the branch from the leaves up to site $i$
\begin{equation}
\left| b_i^{\alpha}\right\rangle = \sum_{b_j^1 \ldots b_j^{z-1} n_{j}}
A_{b_j^1 \ldots b_j^{z-1} b_i^{\alpha}}^{n_j} \left|b_j^1 \ldots b_j^{z-1} n_{j} \right\rangle\ ,
\end{equation}
where the sites $j$ are adjacent to $i$ in the branch\cite{naka2013}. The absence of 
loops in the tree graph simplifies many mathematical properties of the TTNSs and makes them 
similar to MPSs\cite{naka2013}. This allows Nakatani and Chan to use the DMRG optimization algorithm for TTNSs, 
where one site at a time is considered. 

TTNSs approximate many-dimensional entanglement 
by a tree-entanglement structure, which can still be inappropriate for molecules 
with some extended two- and three-dimensional structure, give nonuniform entanglement, and lead to 
convergence problems. In general, molecular orbitals (even specially prepared ones) are delocalized over more than one atom, and hence 
they may not strictly follow the graph underlying a TTNS. 

By contrast, the Complete Graph Tensor Network States (CGTNS) {\it ansatz}\cite{mart2010b} considers entanglement of {\it all} orbitals 
on equal footing by so-called correlators. The CGTNS approach factorizes the high-dimensional CI
coefficient tensor $C_{n_1 n_2 \ldots n_L}$ into a product of all possible 2-site correlators $C_{n_i n_j}^{[ij]}$
\begin{equation}\label{eq:CGTNS}
\left| \Psi^{2s}_{\rm CGTNS} \right\rangle =\sum_{n_1 n_2 \ldots n_L} \prod_{i \le j} C_{n_i n_j}^{[ij]} 
\left| n_1 n_2 \ldots n_L \right\rangle\ .
\end{equation}
The total number of correlators used in this \textit{ansatz} is equal to $L(L+1)/2$, which makes 
the number of variational parameters in this \textit{ansatz}\ equal to $L(L+1)q^2/2$, where $q$ is the number of local states
($q=2$ for spin orbitals and $q=4$ for spatial orbitals).
One can consider the CGTNS \textit{ansatz} as a generalization of the Correlator Product States \textit{ansatz}  
suggested by Changlani et al.\cite{chan2009}, where correlators were only used between nearest-neighbor 
sites. Higher accuracy can be achieved by invoking higher-order correlators 
(3-site correlators, 4-site correlators, and so forth)\cite{chan2009,mart2010b,kovy2015}. 
While we continue to investigate such general decompositions of CI coefficients \cite{kovy2015}, 
the advantage of the MPS \textit{ansatz} is that it can be efficiently optimized by the DMRG algorithm.

\section{Relativistic Hamiltonians and Symmetries}\label{sec:relH}

In 1928 C.\ G.\ Darwin wrote \cite{darwin1928}: \textit{In a recent paper Dirac has brilliantly
removed the defects before existing in the mechanics of the electron, and has shown 
how the phenomena usually called the ``spinning electron" fit into place in the complete theory.}
Since the non-relativistic Schr{\"o}dinger
equation was spin-free, it was clear at that time that a new formalism was needed to combine quantum theory with Einstein's theory of special relativity.
Since the 1970s numerous unusual features have been recognized in heavy-element
chemistry and spectroscopy that can only be explained considering a relativistic quantum description of electrons
\cite{pyykkoe78,key-1,pyykkoe88,key-2}. The liquid state of mercury under ambient condition \cite{mercury_liquid_2013} and the 
lead battery in cars \cite{key-5} are prominent examples for which these so-called '\textit{relativistic effects}' are in operation.
		
Today, relativistic electronic structure theory is a mature and well-understood field \cite{key-2,key-3}. Once a relativistic Hamiltonian is
chosen, established electronic-structure methods can be employed to approximate the wave function. 
Our {\textsc{QCMaquis}} program package can handle the symmetry properties of the Dirac-Coulomb and Dirac-Coulomb-Breit Hamiltonians 
as well as of their two-component analogs \cite{key-6}. Whereas the first such implementation into a traditional DMRG program \cite{knecht2014} 
could only handle real double groups (\texttt{DG}), these limitations are overcome in {\textsc{QCMaquis}}.
		
In the molecular spinor basis significant computational
savings can be achieved by adopting the symmetries 
that are obeyed by two- and four-component Hamiltonians. In non-relativistic
quantum chemistry, one only needs to treat space inversions and rotations
because all other symmetries can be generated by a successive application
of these two. 
In the relativistic framework, time and space are tied together to the space-time, and hence the time reversal operator $\hat{\mathcal{K}}$
is to be addressed and double point groups need to be taken into account. 
The effect of $\hat{\mathcal{K}}$ on a wave function
$\Psi(t)$ yields the time-reversed wave function $\bar{\Psi}(-t)$. 
It can be shown\cite{key-3} that the pair $\{\Psi,\hat{\mathcal{K}}\Psi=\bar{\Psi}\}$\ corresponds to a 
doubly-degenerate fermionic state function which is called a \textit{Kramers pair} (loosely speaking, the
relativistic analog of two degenerate non-relativistic $\alpha$- and $\beta$-spin orbitals). 
Performing calculations in a Kramers basis reduces the possible number of two-electron integrals
$\left(ij|kl\right)$ arising from all combinations of unbarred
and barred indices, 16 in total, to only six symmetric non-redundant
integrals \cite{thyssen_diss}. 
		
Double groups are constructed from the direct product of point groups and the 
subgroup $\{E,\bar{E}\}$ where $\bar{E}$ represents a rotation through
$2\pi$ and $E$ a rotation through $4\pi$. 
Double groups are in general non-abelian which gives rise to additional complications for symmetry
multiplications in quantum chemistry programs. If time reversal symmetry
can be considered (\emph{e.g.}, in the absence of an external magnetic
field), it can be shown \cite{key-3} that certain classes of two-electron
integrals are real, complex, or can be excluded \emph{a priori}, because
they are equal to zero. 
Finally, for systems of interests to chemists, the number of
particles is conserved, which implies that the unitary one-dimensional group U(1),
can also be included along with all other symmetries introduced above. 
In our case, with DMRG as the post-Hartree-Fock method of choice, \texttt{U1DG} 
symmetry is employed to decrease the number of many-particle
states for
symmetry reasons. Characters
and multiplication tables for $C_{1}$, $C_{i}$, $C_{2}$, $C_{2h}$, $C_{64}$ and $C_{32h}$\ double groups \cite{key-4}\ were 
implemented in {\textsc{QCMaquis}} \cite{key-6}. 
		
The relativistic DMRG model in {\textsc{QCMaquis}} supports 
double group symmetries in order to assign to every site an irreducible
representation corresponding to the spinor placed there. No assumptions
are made with respect to the spinor basis which can be either a Kramers-restricted
or Kramers-unrestricted basis. In addition, no formal distinction is made between
barred and unbarred spinors but simplifications due to the
selected symmetry may lead to an elimination of certain terms in the
Hamiltonian. 
Finally, no explicit reference of two- or four-component quantities
is made inside {\textsc{QCMaquis}} and the only input data for the calculations are the
pre-computed relativistic one- and two-electron integrals from \textsc{Molcas}\cite{molcas8}, 
\textsc{Dalton}\cite{WCMS:WCMS1172}, 
\textsc{Molpro}\cite{molpro}, 
\textsc{Dirac} \cite{DIRAC14} or \textsc{Bagel} \cite{bagel_ref}.

\section{Set-up, Parameter Dependence, and Convergence Acceleration}\label{sec:pardep}

The ability of DMRG to handle active orbital spaces that are much larger than those of conventional CASSCF approaches
comes with an additional set of mainly technical parameters that 
can affect convergence and accuracy \cite{keller2014}. 
Among these parameters are the ordering of the orbitals as sites 
on the one-dimensional lattice, the number of renormalized states $m$, the number of sweeps, and 
the initialization procedure of the MPS in the first sweep, the
so-called warm-up sweep.
This increase in the number of control parameters is a threat to the routine application of DMRG in standard computational chemistry. 
As our aim is to make DMRG a valuable and reliable tool for computational chemistry, easy usage as well as stable and fast 
convergence are of paramount importance for a 'black-box' set-up of such calculations.

While the number of sweeps required for convergence and the number of renormalized states necessary can be easily controlled, 
the ordering of the orbitals on the lattice and the initialization procedure need more sophisticated ideas.
Both problems were adressed by Legeza and co-workers \cite{solyom2003,reiher2011} by making use of entanglement measures for 
the active orbitals expressed in terms of one- and two-orbital von Neumann entropies \cite{Legeza2004,legeza_prl2006,rissler2006}. 
Especially the mutual information matrix $\fett{I}$, which is a measure for the entanglement of pairs of orbitals,
proved to be a valuable tool in the analysis of MPS wave functions and multi-reference wave functions in general.
For fast convergence of a DMRG calculation, it is essential that highly entangled orbitals 
are close to each other on the lattice. This will be guaranteed if the orbitals are ordered according to the Fiedler 
vector,\cite{fiedler1973,fiedler1975} which is the eigenvector corresponding to the second smallest eigenvalue of the graph 
Laplacian $\fett{L}_g$, defined in this case as
$\fett{L}_g = \fett{D}-\fett{I}$, 
where $\fett{D}$ is a diagonal matrix, $D_{ii} = \sum_j I_{ij}$ ($i$ and $j$ are labels for the orbitals on the lattice, i.e.,
for the orbitals chosen to be in the CAS). 
The Fiedler vector minimizes the cost measure\cite{reiher2011}
\begin{equation}
\omega = \sum_{ij} I_{ij} |i-j|^2 \, .
\end{equation}
This Fiedler ordering significantly improves the convergence and is implemented in {\textsc{QCMaquis}}.

Convergence can further be improved by a suitable guess of the environment states in the 
warm-up sweep. While an obvious choice is to start from an MPS that contains a reference determinant (such as 
the Hartree-Fock determinant or the determinant with the largest weight in configuration-interaction language), it is possible to improve on this by including the \textit{most important determinants} into the initial MPS.
These most important determinants are selected by varying the occupation on those sites that have the highest 
one-orbital entropies. If these determinants are further limited to have a specific excitation level with respect to a reference 
determinant, this initialization procedure invented by Legeza\cite{legeza04,legeza10} is referred to as CI-DEAS.
Calculations starting from a CI-DEAS MPS are less prone to get stuck in local minima and show enhanced convergence behaviour\cite{reiher2011,legeza2014,molcaspaper,stein2016}. 
The specific CI-DEAS procedure available in {\textsc{QCMaquis}} is described elsewhere\cite{molcaspaper}.

Although the maximally possible size for an active space is enlarged by DMRG, the choice of a suitable set of active orbitals is still 
largely a matter of experience.
It has already been pointed out in the context of traditional CASSCF methodologies 
that the selection of orbitals is a non-trivial problem and can lead to qualitatively wrong results \cite{gordon1998,gagliardi2008,roos2011}. 
This problem is in general not solved by the possibility of including more orbitals. 
On the contrary, the distinction between non-dynamically and dynamically correlated orbitals is equally important in DMRG \cite{keller_2015_pulay} 
and requires a separate description of dynamical correlation by means of perturbation theory or short-range DFT (see Section \ref{sec:dyncorr}). 
However, entropy-based entanglement measures can be of valuable help for the assessment of a chosen CAS\cite{Boguslawski2012b}.

This entanglement information can already be obtained from a preliminary calculation performed with a low number of renormalized states 
$m$ \cite{stein2016}. 
Combined with the fact that DMRG is an iterative algorithm that allows one to stop a calculation well before full (energy) convergence is reached,
this enables us to quickly assess automatically constructed CASs\cite{stein2016}.
Moreover, such unconverged DMRG calculations can additionally be used for the optimized Fiedler ordering and the CI-DEAS 
initialization procedure\cite{stein2016}, all at low additional cost.

\section{Dynamical Correlation}\label{sec:dyncorr}

Whereas static electron correlation effects can be well described by DMRG on the basis of sufficiently large active orbital spaces,
a remaining, yet essential part of electron correlation, commonly referred to as \textit{dynamical}\ correlation, cannot be accounted for. 
It originates from electronic interactions described between orbitals in the active space and external (inactive and secondary) orbitals 
as well as among inactive and secondary themselves. For quantitative results, it is mandatory to account for dynamical electron correlation.

Following the developments of multi-reference wave functions based on traditional multi-configurational wave function approaches 
such as CASSCF, internally-contracted multireference CI (MRCI) \cite{saitow2013}\ and multireference perturbation theory (MRPT) approaches 
--- most importantly, $N$-electron valence perturbation theory to second-order (NEVPT2) \cite{nevpt-qcmaquis} and complete-active space perturbation theory 
to second order (CASPT2) \cite{kurashige:094104,kurashige2014-cupt2} --- were combined with MPS reference wave functions.

The central idea of all of these methods is to describe dominating static correlation effects by a zeroth-order Hamiltonian,
while capturing dynamical correlation in a subsequent step which follows a 'diagonalize-then-pertub' \cite{shavitt2002}\ strategy.  
The price to pay is the need to calculate $n$-particle reduced 
density matrices ($n$-RDMs, with $n >$ 2 and up to 5) of the DMRG wave function and to carry out a four-index transformation of
all two-electron integrals in the full molecular orbital basis. 
The computational cost of the former scales in a na{\"i}ve implementation approximately as $L^{2n}$\ where $L$\ 
is the number of orbitals defining the active orbital space. 
A possible solution to this problem comprises a cumulant-based reconstruction scheme of higher-order $n$-RDMs, typically for $n=3,4$, 
from the knowledge of the 2-RDM alone (see Ref.\ \citenum{harris2002} for a comprehensive review). 
Although such a reconstruction is appealing, neglecting higher-order cumulants (required for the desired computational savings) 
results in a loss of the $N$-representability of the high-order RDM (meaning that the trace of the matrix does not yield 
the number of active electrons $N$). 
The latter can in turn introduce unphysical solutions to the eigenvalue problem under consideration \cite{saitow2013,kurashige2014-cupt2}. 
For these reasons, our current NEVPT2\cite{nevpt-qcmaquis}$^{\rm c}$and CASPT2 implementations\cite{caspt2-qcmaquis} in {\textsc{QCMaquis}} avoid cumulant approximations, 
although the full elegance of our MPO DMRG program for an efficient calculation of 
3- and 4-RDMs has not been fully exploited yet.

In addition to perturbation theory based methods, we also implemented 
a conceptually different approach based on short-range (sr) DFT\cite{savinbook,savin1995}
that (i) does not require the evaluation of higher-order $n$-RDMs, 
(ii) is capable of simultaneously handling dynamic and static correlation, and (iii) 
combines wave function theory with DFT. 
As such, our DMRG-srDFT approach\cite{hedegaard2015b} preserves all efficiency advantages of DMRG.

Hybrid methods between DFT and wave function theory often face the so-called 'double-counting problem' of 
electron-correlation effects because of the correlation energy functional that introduces correlation effects in
a way unrelated to the multi-determinant \textit{ansatz} for the wave function. This issue can be solved elegantly with a 
range-separation \textit{ansatz} \cite{savinbook,savin1995} where the two-electron 
repulsion operator is separated into a short-range and a long-range part. While such an \textit{ansatz} was 
explored for standard wave function methods\cite{savin1995,savinbook,fromager2007,werner_review10,fromager2010,fromager2013,fromager2015}, 
we introduced the DMRG--srDFT approach \cite{hedegaard2015b}\ where 
long-range electron correlation is treated by MPSs in {\textsc{QCMaquis}}
complemented with a short-range DFT description of the two-electron interaction in \textsc{Dalton} \cite{WCMS:WCMS1172}.  

In contrast to two-step approaches (\textit{vide supra}), the overall scaling of DMRG-srDFT does not exceed that 
of a DMRG calculation since it requires at maximum an additional evaluation of a 1-RDM. 
This feature is particularly advantageous for transition-metal complexes or large organic chromophores when combined, for example, 
with the embedding methods described in Section \ref{sec:emb}.   
Although DMRG-srDFT can, in its present formulation, only be used for state-specific optimization of excited states, 
a simultaneous state-average optimization of ground- and excited states is possible in an ensemble DFT \textit{ansatz} (see, for example, 
Ref.~\citenum{senjean2015}). We are currently exploring the latter option based on a (long-range) DMRG wave function in our laboratory.  
The efficiency of srDFT originates from the description of the Coulomb hole of the electron-electron interaction.
However, this does not account for long-range dynamical correlation effects, which are neglected.

\section{Embedding in a Structured Environment}\label{sec:emb}

In appreciation of the fact that the majority of experimental investigations are carried out in some medium, such as a solvent or a protein environment, 
the {\textsc{QCMaquis}} program is coupled to schemes that can describe such a surrounding environment.
We first focus on the coupling of DMRG to the frozen density embedding scheme (FDE)\cite{cortona1991,wesolowski1993,Jacob2014} 
in {\textsc{QCMaquis}}\cite{dresselhaus2015}.  

The FDE scheme belongs to a group of sub-system approaches in which the total system is partitioned into smaller fragments, thereby reducing 
the total computational cost. 
Density-based sub-system approaches assume that the total density can be described as a sum of the densities of the individual subsystems,
for instance,
\begin{equation}
\rho(\mathbf{r}) = \rho_{\rm act}(\mathbf{r}) + \rho_{\rm env}(\mathbf{r}) , \label{total_density}
\end{equation}
where the environment density $\rho_{\rm env}(\mathbf{r})$ itself can be described as a sum of densities of individual fragments, 
$\rho_{\rm env}(\mathbf{r}) = \sum\limits_{J}\rho^{J}_{\rm env}(\mathbf{r})$.  
The FDE scheme is typically employed as a \textit{focused} model which has shown to be a very successful route to model local chemical phenomena 
such as a solute in a solvent or a chromophore within a protein \cite{Jacob2014}. Traditionally, focused models employ a 
quantum mechanical (QM) method for a pre-defined 
\textit{active region} and a more approximate model for the environment. Some models treat the environment as a structureless continuum, whereas others 
(such as FDE) use 
an explicit description. An example of the former is a polarizable continuum model\cite{tomasi2005}, while the most renown explicit model is 
probably the quantum mechanics / molecular mechanics (QM/MM) coupling scheme\cite{warshel1976,senn2009}. FDE strives for higher accuracy 
than QM/MM 
by moving beyond a purely classical description for the environment 
and by also allowing for a polarization of the environment. 

The FDE approach was originally devised within DFT and thus $\rho_{\rm env}(\mathbf{r})$ is obtained from 
Kohn-Sham DFT calculations of the individual fragments constituting the environment. 
This density is then used to construct an  effective embedding operator which enters the Kohn-Sham equations of the active region, 
thereby including the effect of the surrounding environment. Originally, the environment density was kept frozen which can be a severe approximation in cases of 
large mutual polarization of active region and environment. To handle such cases, the two regions can 
be allowed to polarize each other by iteratively exchanging the role of active and environment sub-systems until convergence, 
known as 'freeze-and-thaw' 
cycles\cite{wesolowski1996}.  
The DMRG-FDE implementation builds upon an extension of the original DFT-in-DFT based scheme in order to 
treat the active region with a wave function method\cite{govind1998,govind1999}.  
The DMRG-FDE electronic energy reads\cite{dresselhaus2015}
\begin{equation}
E_{\rm tot} = E^{\rm DMRG}_{\rm act} + E^{\rm KS-DFT}_{\rm env} + E^{\rm OF-DFT}_{\rm int} , \label{FDE_energy} 
\end{equation}
where the first term is
\begin{align}
E^{\rm DMRG}_{\rm act}= \langle\Psi_{\rm MPS}\vert\hat{H}_{\rm MPO}\vert\Psi_{\rm MPS}\rangle . \label{E_act}
\end{align}
In practice, this term is evaluated as the pseudo-energy
\begin{align}
\mathcal{E}^{\rm DMRG}_{\rm act}= \left\langle\Psi_{\rm MPS}\left\vert\hat{H}_{\rm MPO} 
+ \sum^{N}_{i=1}\hat{v}^{\rm act}_{\rm emb}[\rho_{\rm act},\rho_{\rm env}](\mathbf{r}_i )\right\vert\Psi_{\rm MPS}\right\rangle . \label{pseudo_energy}
\end{align}
which can be optimized in an MPO-based formalism in a way that ensures that the embedding potential is obtained self-consistently\cite{dresselhaus2015}. 
The sum in Eq.~(\ref{pseudo_energy}) runs over the active electrons $N$ and the MPS obtained from 
Eq.~(\ref{pseudo_energy}) yields
the energy in Eq.~(\ref{E_act}). $E^{\rm KS-DFT}_{\rm env}$ in Eq.~(\ref{FDE_energy}) is the environment energy evaluated within DFT, while 
$ E^{\rm OF-DFT}_{\rm int}$ is the interaction between active and environment sub-systems. 
$E^{\rm OF-DFT}_{\rm int} $ also contains  
a so-called non-additive energy correction, arising from the exchange-correlation functional and the kinetic energy operator. 
This non-additive kinetic energy is most efficiently evaluated by orbital-free DFT (OF-DFT), which 
requires accurate orbital-free kinetic energy functionals (see, e.g., ~Ref.~\citenum{Jacob2014} and references cited therein). 

Several other embedding schemes also build on a divide-and-conquer approach. Many of these schemes have complementary strengths and weaknesses. 
In addition to the FDE scheme, we have very recently combined\cite{hedegaard2016} {\textsc{QCMaquis}} with a polarizable embedding (PE) scheme\cite{olsen2010} 
that was shown\cite{hedegaard2015a} to yield promising results in combination with multireference methods.
In PE, environment fragment densities $\rho^{J}_{\rm env}(\mathbf{r})$ are represented by a classical multipole expansion with atom-centered multipoles and
(anisotropic) polarizabilitites.

\section{Structure optimization}\label{sec:gradients}

A major area of research in computational chemistry encompasses the study and prediction of (photo-)chemical reaction mechanisms, 
in which the determination of stable intermediates, the location of transition states and the exploration of excited state reaction pathways 
are crucial tasks that necessitate 
access to a reliable potential energy surface (PES) at hand.
Such stationary states can be determined by calculating the first ('gradient') and second ('Hessian') derivatives 
of the electronic energy with respect to all nuclear displacements at a fixed reference geometry, which 
makes a fast and computationally stable evaluation of gradients and Hessian elements an essential feature of \textit{ab initio} methods. 

Since the majority of structure optimization algorithms follow a gradient-only driven optimization scheme (with approximate Hessian evaluations) 
to find extremal points on the PES, we focus here on the derivative of a DMRG state $\left|\Psi\right\rangle$\ in its orbital-optimized form from a 
DMRG-SCF calculation. 
In this case, the Hellmann-Feynman theorem holds such that no coupled-perturbed equation for the orbital relaxation part needs 
to be solved \cite{shepard2012, hu2015excited}. 
An analytic energy gradient is readily obtained by taking the first derivative of the electronic energy $E^{\rm DMRG}$
with respect to the nuclear displacement vector $\mathbf{R}$\ at a given reference geometry ($\mathbf{0}$),
\begin{equation}\label{eq:gradientss}
\left.\frac{\partial E^{\rm DMRG-SCF}}{\partial \mathbf{R}}\right|_{\mathbf{0}} = \left\langle \Psi_{\rm MPS}  \left| 
\left.\frac{\partial \hat{H}_{\rm MPO}}{\partial \mathbf{R}} \right|_{\mathbf{0}} \right| \Psi_{\rm MPS}\right\rangle\ , 
\end{equation}
where $\hat{H}_{\rm MPO}$\ is the electronic Hamiltonian in MPO format. 
It can then be shown \cite{molcaspaper} that the gradient evaluation in Eq.\ (\ref{eq:gradientss}) reduces to a simple evaluation of 1- and 2-RDMs 
that are to be contracted with the derivatives of the one- and two-electron integrals in full analogy to CASSCF \cite{shepard2012}.

Although a majority of chemical reactions takes place on a single Born-Oppenheimer PES (adiabatic processes), in particular photo-chemical 
processes proceed through one or several conical intersections of two PES along a reaction pathway. Since the  Born-Oppenheimer approximation 
breaks down in the vicinity of a conical intersection, non-adiabatic transitions (driven by non-zero non-adiabatic coupling elements between 
the intersecting states)\ are possible.  
A computationally sound description of such a case is therefore best achieved in a state-averaged wave function optimization approach 
which allows one to treat a number of (near-degenerate) electronic states simultaneously on equal footing. 

In contrast to the state-specific case, where the wave function (and therefore the energy) is fully variational, 
this is no longer the case for the energy of a given target state in a state-averaged space of all states under consideration. 
The gradient evaluation for a target state requires therefore the use of the Lagrange technique \cite{staalring2001analytical, delcey2015analytical}, 
in which the wave function of the target state is further relaxed with respect to all variational parameters (orbital rotations and CI coefficient changes) 
in order to obtain a now fully variational ('state-specific') wave function. With the latter at hand, the gradient of the target state in a 
state-average wave function optimization can then be evaluated according to Eq.~(\ref{eq:gradientss})  \cite{molcaspaper}.  

Within {\textsc{QCMaquis}}, a target state can easily be tracked during the structure optimization by using the MPS of the preceding step as a 
starting guess for the current step\cite{keller2015efficient,molcaspaper}. This procedure ensures a maximum overlap between both states. Additionally, each state 
is successively calculated \cite{keller2015efficient} which prevents state-flipping or state-crossing in case of near degeneracies.
The latter task is less trivial within the framework of traditional DMRG, because in this case all states are calculated simultaneously. 
If state-flipping or crossing occurs, it will be more difficult to track the target root for requiring particular tools such as the maximum overlap 
technique \cite{hu2015excited}.

\section{The Singlet-Triplet Gap of Methylene}\label{sec:methylene}

As an example, we present results for the singlet-triplet (S-T) gap of methylene,
CH$_2$. Although CH$_2$ is small and clearly not a typical target for DMRG-based approaches, it 
is a benchmark molecule for new theoretical methods \cite{taylor_fci_ch2}. 
Here, we select small active spaces to demonstrate that DMRG-SCF (with NEVPT2) yields the same
results as traditional CASSCF-type approaches.
Methylene has, in accordance with Walsh's rules, a bent $C_{2v}$\ equilibrium structure (see
Ref.~\citenum{staemmler1973}\ for a qualitative study on the angle
dependence of the S-T gap). 
The HOMO of the singlet $\tilde{a}^1A_{1}$\ state is doubly occupied and of 
symmetry $a_1$. 
In the triplet $\tilde{X}^3B_{1}$\ state
this electron pair becomes
unpaired with one electron now residing in an orbital of
symmetry $b_{1}$
that corresponds to the LUMO in the singlet state.

Unlike its heavier valence-isoelectronic homologs silylene (SiH$_2$)
and germylene (GeH$_2$), which feature ground states of singlet spin symmetry \cite{sih2_exp,geh2_exp},
methylene has a triplet degenerate ground state with the lowest-energy singlet state lying about 9.2
kcal/mol (9.0 kcal/mol including zero-point vibrational corrections)
higher in energy \cite{ch2_exp}. A qualitative explanation for this
observation could be based on the magnitude of the HOMO-LUMO gap which increases
from CH$_2$ to SiH$_2$\ to GeH$_2$, but as discussed in 
detail in Ref.~\citenum{apeloig2003}\ other electronic and steric
effects need to be taken into account to arrive at a
quantitative understanding.

For the results presented in this work, we employed a cc-pVTZ basis set
\cite{dunning89}. The equilibrium structures of the triplet
$\tilde{X}^3B_{1}$\ ground state and the first excited singlet
$\tilde{a}^1A_{1}$\ state correspond to those determined
by Sherrill and co-workers with an FCI/TZ2P approach
\cite{fci_tz2p_ch2}. Adapting a $C_{2v}$\ structure, the
HCH angle and C-H bond lengths are 133.29$^{\circ}$\
(101.89$^{\circ}$) and 1.0775 \AA\ (1.1089 \AA), respectively, in the
triplet (singlet) state.

We carried out a series of state-specific CASSCF, CASSCF/CASPT2, DMRG-SCF, and
DMRG-SCF/NEVPT2 calculations with increasing size of the active orbital
space to study the S-T splitting in methylene. The CASSCF and
CASSCF/CASPT2 calculations were performed with a developers' version of
the \textsc{Molcas 8}\ \cite{molcas8}\ software package with its
default zeroth-order Hamiltonian for CASPT2.
The DMRG-SCF and DMRG-SCF/NEVPT2 calculations were carried out with 
{\textsc{QCMaquis}} and our
local NEVPT2 implementation \cite{nevpt-qcmaquis}$^{\rm c}$. For DMRG-SCF/NEVPT2
we report only data for the so-called partially contracted approach
as the results for the strongly contracted approach are similar.
The number of renormalized DMRG block states $m$\ was set to $m$=1024
which is sufficient to reach CASSCF accuracy for those active orbital spaces
where a comparison with traditional CASSCF data was possible.
Our CAS(6,6) comprises three orbitals in symmetry $a_1$, one
in  $b_1$, and two in $b_2$, while the 
CAS(6,12) comprises six orbitals in $a_1$, two in $b_1$, and four in $b_2$.
%
%
For comparison, we also performed single-point DFT calculations with the 
PBE\cite{perdew1996b}
and
PBE0\cite{perdew1996a}
density functionals (as implemented in the Turbomole 6.5 program package \cite{Ahlrichs1989}).
All results are compiled in Table \ref{tab:ch2} for the S-T splitting in
methylene together with previous theoretical results and the experimental
reference value.

\begin{table}[H]
     \begin{center}
     \caption{Calculated adiabatic singlet-triplet gap,
$E(\tilde{a}^1A_{1})$--$E(\tilde{X}^3B_{1})$, in kcal/mol for methylene.
\label{tab:ch2}}
     \begin{tabular}{lccc}
         \hline \hline \\
       Method &  \multicolumn{2}{c}{singlet-triplet gap}\\ \hline
                & CAS(6,6) & CAS(6,12) & CAS(6,20) \\
         CASSCF & 10.53    &  5.71      &  9.93  \\ 
         CASSCF/CASPT2& 11.87&10.56     &  10.26   \\ 
         DMRG-SCF & 10.53    &5.71      &  9.93    \\ 
         DMRG-SCF/NEVPT2 & 11.71&9.13   &  10.17         \\ 
       \hline
         PBE & \multicolumn{2}{c}{16.03}\\
         PBE0 & \multicolumn{2}{c}{17.72}\\
       \hline
                           &     \multicolumn{2}{c}{previous work} \\ 
         CAS-BCCC4 (Ref.~\citenum{block_cc_in_book2010}) & \multicolumn{3}{c}{9.60} \\
         MR--CISD+Q (Ref.~\citenum{block_cc_in_book2010}) & \multicolumn{3}{c}{9.68} \\
         FCI$^a$ (Ref.~\citenum{fci_tz2p_ch2}) & \multicolumn{3}{c}{11.14}\\
         MR(6,12)MP2$^b$ (Ref.~\citenum{hirao_rev2005}) & \multicolumn{3}{c}{9.9}\\
         CCSDT (Ref.~\citenum{bartlett2014_ch2})& \multicolumn{3}{c}{9.0$^c$}\\
         Exp.\ (Ref.~\citenum{ch2_exp}) & \multicolumn{3}{c}{8.99$^d$/9.37$^d$}\\
         \hline \hline
         \multicolumn{4}{l}{$^a$\ TZ2P basis set; one core and one virtual orbital frozen.}\\
         \multicolumn{4}{l}{$^b$\ cc-pVTZ basis set; equilibrium structures taken from Ref.~\citenum{taylor_fci_ch2}.}\\
         \multicolumn{4}{l}{$^c$\ Equilibrium structures optimized with CCSD(T)/6- 311++G(2d,2p);}\\
         \multicolumn{4}{l}{extrapolation to the complete basis set limit from CCSDT/cc-pVTZ and}\\
         \multicolumn{4}{l}{CCSDT/cc-pVQZ calculations.}\\
         \multicolumn{4}{l}{$^d$\ Modified for direct comparison with the electronic energy difference;}\\
         \multicolumn{4}{l}{see also Refs.~\citenum{ch2_exp,bartlett2014_ch2}.}\\
     \end{tabular}
         \end{center}
\end{table}

As expected, the deviations of our S-T splittings from the
experimental reference decrease with an increasing active orbital space
(with the exception of CAS(6,12) where clearly a poor reference yields accidentally a seemingly accurate 
DMRG-SCF/NEVPT2 result). 
Moreover, our calculated S-T splittings agree well with previous theoretical results
obtained by various methods. 
Somewhat surprising is the excellent performance of the single-reference CCSDT model
\cite{bartlett2014_ch2}\ for the multi-configurational character of
the singlet $\tilde{a}^1A_{1}$\ state that is best described in a
two-configuration model \cite{taylor_fci_ch2}. However, our
CASSCF calculation with a CAS(6,20) yields a distribution of 91\%
of configuration $2a_1^2\ 1b_2^2\ 3a_1^2$\ and 3\% of $2a_1^2\ 1b_2^2\ 1b_1^2$. 
Interestingly, CASPT2 (NEVPT2) does not improve on the CASSCF (DMRG-SCF) S-T splitting
(again with the CAS(6,12) result for DMRG-SCF/NEVPT2 as an exception), but yields a result that
deviates more from experiment. This observation was already made for CASPT2
in Ref.\ \citenum{block_cc_in_book2010}. In addition, we also observe a
similar trend for DMRG-SCF/NEVPT2.  A possible explanation might be a
differential dynamical-correlation effect where less dynamical correlation
is recovered for the singlet $\tilde{a}^1A_{1}$\ state than for the triplet state by either variant
of multi-reference perturbation theory that in turn leads to an
overstabilization of the triplet $\tilde{X}^3B_{1}$\ ground state.
Finally, note that the DFT S-T gaps are off by almost a factor of two.

\section{Conclusions}\label{sec:conclusions}

A reliable computational exploration of complex chemical reactions requires
sophisticated \textit{ab initio}\ approaches that are capable of accurately 
describing an electronic structure dominated by strong static correlation. The quantum-chemical DMRG algorithm 
iteratively converges to the exact solution of the electronic Schr{\"o}dinger equation within a given complete active (orbital) 
space. Unlike traditional CAS-based approaches, which suffer from an exponential scaling of the computational cost with respect to an 
increase in the number of active orbitals and electrons, 
DMRG with its polynomial scaling emerged as a new option to explore the spectroscopy and chemical reactivity of molecular complexes 
with active spaces comprising up to 100 orbitals. Different possibilities are available to take into account dynamical correlation effects 
based on a DMRG wave function. Structure optimizations in ground and electronically excited states are possible. Also an embedding 
in a structured environment can be efficiently modeled. Finally, relativistic DMRG models allow us to account for spin-orbit coupling 
and other 'relativistic effects' in a rigorous way. 
All these building blocks complete our new computational toolbox {\textsc{QCMaquis}} that is available free of charge from our webpage\cite{reiherwebsite}.

\section*{Acknowledgments}

This work was supported by ETH Zurich (Research Grant ETH-34 12-2 and ETH Fellowship FEL-27 14-1) 
and by the Schweizerischer Nationalfonds (Project No. 200020\_156598).
CJS thanks the Fonds der Chemischen Industrie for a K{\'e}kule PhD fellowship.
EDH thanks the Villum foundation for a post-doctoral fellowship.


\footnotesize
\newcommand{\Aa}[0]{Aa}
\providecommand{\url}[1]{\texttt{#1}}
\providecommand{\urlprefix}{}
\providecommand{\foreignlanguage}[2]{#2}
\providecommand{\Capitalize}[1]{\uppercase{#1}}
\providecommand{\capitalize}[1]{\expandafter\Capitalize#1}
\providecommand{\bibliographycite}[1]{\cite{#1}}
\providecommand{\bbland}{and}
\providecommand{\bblchap}{chap.}
\providecommand{\bblchapter}{chapter}
\providecommand{\bbletal}{et~al.}
\providecommand{\bbleditors}{editors}
\providecommand{\bbleds}{eds.}
\providecommand{\bbleditor}{editor}
\providecommand{\bbled}{ed.}
\providecommand{\bbledition}{edition}
\providecommand{\bbledn}{ed.}
\providecommand{\bbleidp}{page}
\providecommand{\bbleidpp}{pages}
\providecommand{\bblerratum}{erratum}
\providecommand{\bblin}{in}
\providecommand{\bblmthesis}{Master's thesis}
\providecommand{\bblno}{no.}
\providecommand{\bblnumber}{number}
\providecommand{\bblof}{of}
\providecommand{\bblpage}{page}
\providecommand{\bblpages}{pages}
\providecommand{\bblp}{p}
\providecommand{\bblphdthesis}{Ph.D. thesis}
\providecommand{\bblpp}{pp}
\providecommand{\bbltechrep}{Tech. Rep.}
\providecommand{\bbltechreport}{Technical Report}
\providecommand{\bblvolume}{volume}
\providecommand{\bblvol}{Vol.}
\providecommand{\bbljan}{January}
\providecommand{\bblfeb}{February}
\providecommand{\bblmar}{March}
\providecommand{\bblapr}{April}
\providecommand{\bblmay}{May}
\providecommand{\bbljun}{June}
\providecommand{\bbljul}{July}
\providecommand{\bblaug}{August}
\providecommand{\bblsep}{September}
\providecommand{\bbloct}{October}
\providecommand{\bblnov}{November}
\providecommand{\bbldec}{December}
\providecommand{\bblfirst}{First}
\providecommand{\bblfirsto}{1st}
\providecommand{\bblsecond}{Second}
\providecommand{\bblsecondo}{2nd}
\providecommand{\bblthird}{Third}
\providecommand{\bblthirdo}{3rd}
\providecommand{\bblfourth}{Fourth}
\providecommand{\bblfourtho}{4th}
\providecommand{\bblfifth}{Fifth}
\providecommand{\bblfiftho}{5th}
\providecommand{\bblst}{st}
\providecommand{\bblnd}{nd}
\providecommand{\bblrd}{rd}
\providecommand{\bblth}{th}

\end{document}